\documentclass[sigconf]{acmart}

\acmConference[ICSE 2024]{46th International Conference on Software Engineering}{April 2024}{Lisbon, Portugal}
\renewcommand\footnotetextcopyrightpermission[1]{}
\usepackage{graphicx}

\usepackage{listings}
\usepackage{tikz}
\usetikzlibrary{tikzmark,calc}
\usepackage{color}
\usepackage{xcolor}
 \usepackage{enumitem}

\AtBeginDocument{%
  \providecommand\BibTeX{{%
    \normalfont B\kern-0.5em{\scshape i\kern-0.25em b}\kern-0.8em\TeX}}}

\usepackage{multirow}
\usepackage{tabularx}

\begin{document}

\title{Exploring the Potential of ChatGPT in Automated Code Refinement: An Empirical Study}



\author{Qi Guo}
\authornote{This work was done while both authors were visiting students at Singapore Management University. Both authors contributed equally to this work.}
\affiliation{%
  \institution{Tianjin University}
  \city{Tianjin} 
  \country{China}} 

\author{Junming Cao}
\authornotemark[1]
\affiliation{%
  \institution{Fudan University}
  \city{Shanghai} 
  \country{China}} 

\author{Xiaofei Xie}
\affiliation{%
  \institution{Singapore Management University}
  \country{Singapore}}

\author{Shangqing Liu}
\authornote{Corresponding author.}
\affiliation{%
  \institution{Nanyang Technological University}
  \country{Singapore}}

\author{Xiaohong Li}
\authornotemark[2]
\affiliation{%
  \institution{Tianjin University}
  \city{Tianjin} 
  \country{China}} 

\author{Bihuan Chen}
\affiliation{%
  \institution{Fudan University}
  \city{Shanghai} 
  \country{China}} 

\author{Xin Peng}
\affiliation{%
  \institution{Fudan University}
  \city{Shanghai} 
  \country{China}} 
\renewcommand{\shortauthors}{Guo and Cao, et al.}
\newcommand{\cao} [1]{\textcolor{black}{{\sf Cao}: #1}}
\newcommand{\todo}[1]{\textcolor{black}{#1}}
\newcommand{\sql}[1]{\textcolor{black}{sql: #1}}
\newcommand{\guo}[1]{\textcolor{black}{#1}}
\newcommand{\cam}[1]{\textcolor{black}{#1}}
\begin{abstract}
  Code review is an essential activity for ensuring the quality and maintainability of software projects. However, it is a time-consuming and often error-prone task that can significantly impact the development process. Recently, ChatGPT, a cutting-edge language model, has demonstrated impressive performance in various natural language processing tasks, suggesting its potential to automate code review processes. However, it is still unclear how well ChatGPT performs in code review tasks. To fill this gap, in this paper, we conduct the first empirical study to understand the capabilities of ChatGPT in code review tasks, specifically focusing on automated code refinement based on given code reviews. To conduct the study, we select the existing benchmark CodeReview and construct a new code review dataset with high quality. We use CodeReviewer, a state-of-the-art code review tool, as a baseline for comparison with ChatGPT. Our results show that ChatGPT outperforms CodeReviewer in code refinement tasks. Specifically, our results show that ChatGPT achieves higher EM and BLEU scores of 22.78 and 76.44 respectively, while the state-of-the-art method achieves only 15.50 and 62.88 on a high-quality code review dataset.  We further identify the root causes for ChatGPT's underperformance and propose several strategies to mitigate these challenges. Our study provides insights into the potential of ChatGPT in automating the code review process, and highlights the potential research directions.

\end{abstract}




\maketitle
\section{Introduction}

Code review is a software quality assurance activity in software development and maintaince, which involves the systematic examination of source code to identify and rectify errors, improve code quality, and ensure compliance with coding standards. The code review process typically consists of writing code reviews and refining code based on the review comments received, with the ultimate goal of enhancing software quality. Code review has become an integral part of many software development projects, as it has been widely recognized for its effectiveness in improving the overall reliability and maintainability of software systems.

However, code review can be a time-consuming and resource-intensive process, requiring significant manual effort to review and refine code, especially in popular projects with numerous contributions. For example, Bosu et al.~\cite{bosu2013impact} discovered that, on average, developers allocate approximately six hours per week preparing code for review or reviewing others’ code.
Moreover, the increasing complexity of modern software systems and the need for more frequent releases have made code review even more challenging. To address this issue, recent research~\cite{thongtanunam2022autotransform,tufano2019learning} has been conducted to automate various aspects of code review, such as generating review comments and refining code. In particular, the learning-based approaches~\cite{ms2022codereviewer,tufano2021towards} that rely on Large Language Models (LLMs) such as CodeT5~\cite{codet5} and CodeBERT~\cite{feng2020codebert} have demonstrated promising results in automating code review, reducing the manual effort required for code reviews.

Recently, OpenAI introduced ChatGPT~\cite{ChatGPTblog}, a revolutionary technology capable of transforming various sectors, including software engineering tasks. ChatGPT, an advanced version of GPT-3.5~\cite{instructgpt}, is a fine-tuned model that excels at understanding and executing instructions. This capability distinguishes it from other pre-trained models and makes it a promising candidate for tasks that require prompts or instructions. The code refinement process, which is contingent upon code review and previous code versions, aligns well with strengths of ChatGPT. Since human reviews can serve as prompts for code refinement, it is natural to investigate the potential of using ChatGPT for this task.

In this paper, we take the first step towards investigating the potential of ChatGPT for code refinement based on the given review comments. \guo{Note that although code-to-code refinement (i.e., ChatGPT directly generates refined code from original code) is also a research problem, there are still major concerns regarding the quality of the refined code~\cite{tufano2022using}. Therefore, we focus on the refinement based on given review in this paper, which is different from code-to-code refinement.}
Specifically, we focus on three main problems: 1) How does ChatGPT perform compared to the state-of-the-art methods? 2) In which cases does ChatGPT underperform, and what are the underlying reasons? 3) How thse challenges can be mitigated? By answering these questions, we can gain a deeper understanding of the potential and challenges of ChatGPT for automated code refinement tasks.

To answer the above questions, we conduct comprehensive experiments to evaluate ChatGPT's performance in code refinement tasks. Considering the sensitivity of ChatGPT to different settings, we first design the experiment to evaluate its performance on two main factors, i.e., different prompts and temperatures. Then we select the optimal configuration and compare ChatGPT with state-of-the-art techniques~\cite{codet5} on standard benchmarks~\cite{ms2022codereviewer}. To evaluate the generalizability of different techniques, we create a new dataset by collecting code reviews from repositories not included in the standard benchmarks and recent code reviews from the same repositories included in the standard benchmarks. Based on the evaluation results, we perform an in-depth analysis of the root causes and devise preliminary strategies for mitigating different challenges.


Overall, the results provide valuable insights into the performance of ChatGPT in code refinement tasks. Our findings demonstrate that different prompts and temperature settings can have a significant impact of up to 5\% and 15\% on ChatGPT's Exact Match (EM) scores in code refinement tasks. Lower temperature settings yield better and more stable results, and describing the code review scenario in the prompt helps enhance ChatGPT's performance. Compared to the state-of-the-art model CodeReviewer, ChatGPT demonstrates better generalization capabilities in our newly generated dataset. Specifically, ChatGPT achieves EM and BLEU scores of 22.78 and 76.44, respectively, on the new dataset, while CodeReviewer only reaches 15.50 and 62.88 for EM and BLEU scores, respectively. However, we also found that ChatGPT struggles on tasks involving refining documentation and functionalities, mainly due to a lack of domain knowledge, unclear location, and unclear changes in the review comments. These limitations could potentially be resolved by improving review quality and using more advanced large language models such as GPT-4. Our study highlights the potential of ChatGPT in code refinement tasks and identifies important directions for future research.


In summary, this paper makes the following contributions:
\begin{itemize}[leftmargin=*]
    \item We conduct the first empirical study on evaluating ChatGPT's potential in code refinement tasks based on review comments.
    \item We analyze the challenges of ChatGPT in code refinement tasks and propose potential mitigation strategies, laying the groundwork for future research on better incorporating ChatGPT.
    \item  We release a new dataset that contains high-quality code reviews,  which could be useful for future research in this area.
\end{itemize}

\section{Background}
\subsection{Code Review Process}
During the code review process, a contributor submits code changes to implement new features, refactor code, or fix bugs.
When the contributor believes the code changes are ready for review and to be merged into the main branch, he or she initiates a pull request and invite reviewers to examine the changes. After reviewing the code changes, a reviewer may provide review comments in natural language, represented as $R$. Based on these review comments, the contributor makes modifications on the original code $C_1$ and submits the revised code $C_2$. The code difference between $C_1$ and $C_2$ is denoted as $D: C_1 \to C_2$.
It is worth noting that the above process represents only one review cycle, while a complete pull request may involve multiple rounds of review cycles. In this work, without loss of generality, we focus solely on the single-round scenario, where to generate the revised submitted code $C_2$ with models automatically, based on the a given review comment $R$ and the original submitted code $C_1$ within each pull request.


\subsection{ChatGPT}

ChatGPT~\cite{ChatGPTblog} is a famous example of Large language models (LLMs), unveiled by OpenAI. 
ChatGPT was developed by employing a GPT-3.5 series model and training it using reinforcement learning from human feedback (RLHF)~\cite{instructgpt,stiennon2020learning}. Owing to the RLHF training process, ChatGPT has exhibited remarkable proficiency across multiple dimensions, encompassing the generation of high-quality responses to human inputs, the refusal of inappropriate queries, and the capacity for self-correction of prior errors based on subsequent dialogues.

Considering the characteristics of ChatGPT usage~\cite{qin2023chatgpt}, it is natural to explore its potential in automating code reviews~\cite{zhou2023generation}. Specifically, we propose a conversational approach to delegate the code refinement task to ChatGPT, where the original code and review comment are provided as a task input in a coherent linguistic structure. ChatGPT will return the revised code along with the reasoning behind the modifications, precisely aligning with the desired output of the task. The performance of ChatGPT in this approach depends significantly on two parameters: prompt and temperature. The prompt serves as a cue for ChatGPT to understand the intended task, while temperature can be used to control the level of creativity and diversity in responses of ChatGPT.

\begin{figure*}[!t]
    \centering
    \includegraphics[scale=0.53]{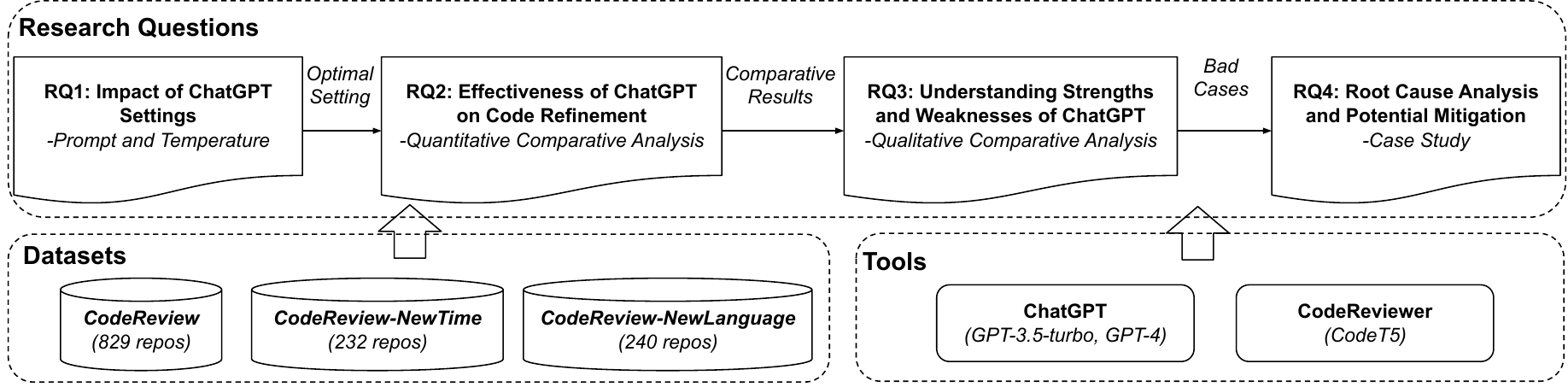}
    \caption{Overview of our study}
    \label{fig:overview}
\end{figure*}

\section{STUDY DESIGN}
\subsection{Overview and Research Questions}
The main focus of this paper is to evaluate and understand the capabilities of ChatGPT in code refinement tasks. Fig.~\ref{fig:overview} shows the overview of this paper. To conduct our study, we collect existing benchmarks, including the CodeReview dataset, and state-of-the-art code refinement tools such as CodeReviewer~\cite{ms2022codereviewer}, for comparisons. However, given the potential risk that the dataset could be used to be trained in ChatGPT and CodeReviewer, we create a new code review dataset (named CodeReview-New) consisting of two parts: new code reviews from the same repositories as CodeReview dataset but collected more recently (i.e., CodeReview-NewTime), and code reviews from repositories using different languages that are not included in CodeReview dataset (i.e., CodeReview-NewLanague). We next introduce the research questions we aim to investigate and their relationships.


{\bfseries RQ1 Impact of ChatGPT Settings: How do different prompt and temperature settings affect ChatGPT's performance in the code refinement task?} As the effectiveness of ChatGPT highly depends on the prompts and temperatures used, we first evaluate the impact of different settings of ChatGPT on code refinement. We designed five prompts based on whether a concreate scene is provided and whether detailed requirements are given. We also selected five temperature settings ranging from 0 to 2, with intervals of 0.5  (i.e., 0, 0.5, 1, 1.5 and 2.0). We evaluated and compared the effects of 25 combinations of these five prompts and five temperature settings based on the CodeReview dataset. Our evaluation of ChatGPT in the subsequent research questions is based on the optimal prompt and temperature settings obtained from RQ1.

{\bfseries RQ2 Effectiveness of ChatGPT on Code Refinement: How does ChatGPT's performance compare to state-of-the-art methods?}
We aim to investigate the effectiveness of ChatGPT in code refinement tasks compared to state-of-the-art methods. To answer this question, we compare ChatGPT's performance with that of the state-of-the-art code refinement tool, CodeReviewer~\cite{ms2022codereviewer}. We replicated and fine-tuned the CodeReviewer model and evaluated its performance alongside ChatGPT on both the existing CodeReview dataset and the new dataset CodeReview-New we created. 

{\bfseries RQ3 Strengths and Weaknesses of ChatGPT: In which cases does ChatGPT perform well or not?}
To address this question, we conduct a qualitative study based on the results obtained from RQ2. Specifically, we annotate 200 samples each from the CodeReview and CodeReview-New datasets manually, labeling the quality of reviews (i.e., relevance and information levels) and code change types. We then evaluate the performance of ChatGPT on data with   various review qualities and code change categories.


{\bfseries RQ4 Root Causes and Potential Mitigation Strategies for Underperforming Cases: What are the underlying causes for the underperformance of ChatGPT, and how can we mitigate these challenges?}
Based on the analysis of RQ3, we aim to further understand the root causes of ChatGPT's underperforming cases and how to address this limitations. We investigated the 206 cases from the 400 annotated samples in RQ3 where ChatGPT failed to make accurate predictions and summarized the categories of root causes. \guo{Based on the root causes, we attempt to study the impact of improving review quality and enhancing models in mitigating the issues of ChatGPT.}

\subsection{Experiment Settings}
\subsubsection{Dataset}
To conduct the study, we utilize two datasets:  the CodeReview dataset ~\cite{ms2022codereviewer} and a new dataset created by us, denoted as CodeReview-New.

\textbf{CodeReview (CR):} 
We first select CodeReview~\cite{ms2022codereviewer}  that is a widely used dataset in code review task. This dataset was crawled from the top 10,000 repositories from GitHub based on their star ranking, and includes nine programming languages, namely C, C++, C\#, Go, Java, JavaScript, PHP, and Python. Repositories that do not have an explicit data redistribution license and fewer than 1,500 pull requests (PRs) are filtered out. The dataset consists of review comments $R$ associated with their corresponding code diff $D: C_1 \to C_2$. To ensure a high-quality dataset, samples with the same review comment associated with multiple code diffs or a single code diff associated with multiple comments are filtered out. Additionally, the dataset is divided into a pre-training dataset and multiple downstream task datasets, and we used the code refinement downstream task dataset in our study. This dataset comprises  829  repositories and  125,653 PRs. \guo{We follow the same partition method as CodeReviewer~\cite{ms2022codereviewer} for a fair comparison, and divide the dataset into training set, validation set and test set, with proportions of 85\%, 7.5\%, and 7.5\%, respectively}. 

\textbf{CodeReview-New (CRN):} Additionally, we create a new code review dataset, CodeReview-New, due to two reasons: 1) we observe that there are some low-quality code review data in CodeReview, which could affect the comparisons between ChatGPT and the baseline CodeReviewer; 2) the data distribution in the CodeReview test data could be very similar to that in  the pre-train and fine-tuning dataset, and may even have been used by the selected models (i.e., ChatGPT~\cite{instructgpt} and CodeT5~\cite{codet5}). The new dataset is constructed to better evaluate the generalization capabilities of models. 
To address these two concerns, we design more strict filtering rules to filter low-quality reviews; and select code reviews that is unlikey to be used in the pre-training process.

To ensure the quality of the CodeReview-New dataset, we implemented several strict rules based on our analysis of the quality issues present in CodeReview. Only code reviews that met these rules were retained in our dataset. Firstly, we ensured that the code changes are only about a single code hunk, which is necessary because the baseline CodeReviewer we select only accepts a single piece of code as input. Secondly, we filtered out changes that were unrelated to code, such as changes to README files. Finally, we ensured the relevance between the review comment $R$ and the code changes $D$ by collecting the original code piece $C_1$ that contains the review comment $R$.

To prevent ChatGPT from using CodeReview-New during the pre-training process, we only collected data from January 1, 2022, onwards, as ChatGPT's training data only extends up to 2021~\cite{ChatGPTblog}. Furthermore, CodeReview dataset also does not contain data after January 1, 2022, which makes it fair to compare CodeReviewer model and ChatGPT. 
In addition to the repositories included in CodeReview, we crawled code reviews from additional 1,400 repositories (top 200 repositories for each language based on their star ranking) using seven programming languages: Swift, Objective-C, Kotlin, SQL, Perl, Scala, and R, which are not included in CodeReview. 
In total, we selected 2,029 repositories, with 829 from the CodeReview repository and 1,200 new repositories with different programming languages.

After applying the filtering rules and selecting pull requests based on time, we only have 467 repositories out of the initial 2,029 repositories. The exclusion of the other 1,562 repositories can be attributed to two main reasons: first, we used stricter filtering rules compared to the construction of the CodeReview dataset, and second, we only selected pull requests created on or after January 1, 2022, which resulted in the exclusion of some projects that had few PRs during this period.
As shown in Table \ref{tab:codereview_new_sta}, the dataset consists of samples from two types of repositories: 9,117 samples from 232 repositories that are also included in the CodeReview dataset, denoted as CodeReview-NewTime (CRNT); 5,451 samples from 240 new repositories that have different programming languages with the repositories in CodeReview dataset, denoted as CodeReview-NewLanguage (CRNL). 
Some languages, such as SQL and Perl, have a smaller amount of data due to fewer pull requests or a smaller number of reviews.


\begin{table*}[!t]
\centering
\footnotesize
\caption{The statistics of CodeReview-New dataset.}
\begin{tabular}{cccccccccccccccccccc}
\hline

\multicolumn{1}{c}{\multirow{2}{*}{Language}} & \multicolumn{1}{c}{} & \multicolumn{9}{c}{$CRNT$}                                                                                                                                                                                                                  & \multicolumn{1}{c}{} & \multicolumn{7}{c}{$CRNL$}                                                                                                                                                      &  \\ \cline{3-11} \cline{13-19}
\multicolumn{1}{c}{}                          & \multicolumn{1}{c}{} & \multicolumn{1}{l}{Ruby} & Go                        & Py                        & C\#                     & JS                      & C++                     & Java                      & C                       & PHP                     &                      & \multicolumn{1}{l}{Swift} & Obj-C                  & Kt                        & SQL                    & PL                      & Scala                     & R                       &  \\ \hline
\#Samples                                     &                      & 377                      & \multicolumn{1}{c}{2,843} & \multicolumn{1}{c}{2,115} & \multicolumn{1}{c}{703} & \multicolumn{1}{c}{427} & \multicolumn{1}{c}{700} & \multicolumn{1}{c}{1,194} & \multicolumn{1}{c}{335} & \multicolumn{1}{c}{423} & \multicolumn{1}{c}{} & 864                       & \multicolumn{1}{c}{81} & \multicolumn{1}{c}{1,932} & \multicolumn{1}{c}{96} & \multicolumn{1}{c}{116} & \multicolumn{1}{c}{1,682} & \multicolumn{1}{c}{680} &  \\ \hline
Total                                         &                      & \multicolumn{9}{c}{9,117}                                                                                                                                                                                                                      &                      & \multicolumn{7}{c}{5,451}                                                                                                                                                               &  \\ \hline
\end{tabular}
\label{tab:codereview_new_sta}
\end{table*}

\subsubsection{Evaluation Models}
To compare the performance of ChatGPT with the state-of-the-art tool, we chose CodeReviewer~\cite{ms2022codereviewer}, which is a recent state-of-the-art method for code refinement. In this paper, we apply ChatGPT in a similar way to CodeReviewer, by generating revised code $C_2$ based on reviews $R$ and original code $C_1$. We chose CodeReviewer over other methods as it is demonstrated to be more effective than other methods such as AutoTransform~\cite{thongtanunam2022autotransform} and Trans-Review~\cite{tufano2021towards}. Based on our evaluation results, we believe that ChatGPT can also surpass other models. Furthermore, our main focus is to understand the strengths and weaknesses of ChatGPT and identify potential improvement directions for future research on the code review process.


\textbf{CodeReviewer:} It utilizes a T5 model architecture comprising 12 Transformer encoder layers and 12 decoder layers, amounting to 223 million parameters ~\cite{t5}. The model is initialized using the weight parameters of CodeT5 ~\cite{codet5}. Subsequently, the pre-training is carried out with three objectives: Diff Tag Prediction, Denoising Objective, and Review Comment Generation. In this study, we employed the same pre-trained CodeReviewer model and fine-tuned it using the $CodeReview_{train}$ and $CodeReview_{valid}$ datasets.

\textbf{ChatGPT:} We accessed and evaluated ChatGPT with the default GPT-3.5-Turbo model using the OpenAI API~\cite{instructgpt}. Unlike CodeReviewer, we did not fine-tune ChatGPT and only performed a zero-shot style evaluation. The ChatGPT API was accessed in March 2023, at a total cost of 150 USD.
When comparing T5 and GPT-3.5, both models are large language models, but they have some differences. T5 is a general-purpose language model that uses a denoising autoencoder objective, which involves predicting masked or corrupted tokens in the input text. In contrast, ChatGPT is trained on a large dataset of conversational text, making it better at generating responses appropriate for use in a chatbot context. One key difference between the two models is that ChatGPT is fine-tuned with Reinforcement Learning from Human Feedback (RLHF), which uses human feedback in the training loop to make it more effective in generating appropriate and coherent responses in various contexts.
During the evaluation, we designed different prompts based on the original code and code review to obtain outputs from ChatGPT. 
In RQ4, we also employed GPT-4 in ChatGPT in order to mitigate the cases where GPT-3.5 made incorrect answers. GPT-4~\cite{openai2023gpt4} is the latest multi-modal model designed to process both textual and visual inputs, generating textual outputs. 



\subsubsection{Evaluation Metrics}
Exact Match (EM) and BLEU are the two widely adopted metrics in previous literature ~\cite{ms2022codereviewer,tufano2021towards,tufano2022using}. However, we found that ChatGPT tends to generate more content including additional code or more explanations, which could largely affect the EM results and make the measurement less accurate. In the real world, a contributor can easily trim these additional information to obtain the correct. 
Hence, we propose two new variants of EM and BLEU, called EM-trim and BLEU-trim, which more accurately measures the results.

\textbf{Exact Match (EM).} A prediction is considered correct by EM only if the predicted revised code is identical to the ground truth revised code. The EM value is computed based on the percentage of generated outputs that exactly match the ground truth.

\textbf{Exact Match Trim (EM-trim)} is a variant of the EM metric that is more lenient in its measurement. EM-trim first performs a trim on the generated output (denoted as $C_2'$) before calculating the EM score. 
Specifically, if the first line of the ground truth text can be found in the generated output $C_2$, we trim the generated content before the first line of $C_2$. Similarly, if the last line of the ground truth text can be found in the generated output $C_2$, we trim the generated content after the last line of $C_2$. After the trim process, the EM-trim score is calculated using the trimmed content $C_2'$ and the ground truth text. The EM-trim metric is more lenient than the traditional EM metric, as it ignore other irrelevant information.


\textbf{BLEU} is a common metric used to measure the quality of generated text in neural translation models~\cite{bleu}. We use the BLEU-4 variant, which calculates the overlap of 4-grams between $C_2$ and the ground truth~\cite{tufano2021towards, ms2022codereviewer, tufano2022using}. The range of BLEU-4 scores lies between 0\% and 100\%, with 100\% indicating a perfect match. The average BLEU-4 score of all test samples serves as the overall evaluation result. Similar to EM-trim,  we also design \textbf{BLEU-trim} that calculates the BLEU-4 score between the trimmed output $C_2'$ and the ground truth text.

\section{Evaluation Results}

\subsection{RQ1 Impact of Prompts and Temperatures}

\begin{table*}[!t]
\caption{Impact of different prompts and temperatures on performance of ChatGPT.}
\centering
\footnotesize 
\label{tb:settingimpact}
\resizebox{1\textwidth}{!}{
\begin{tabular}{ccccccccccccccccccc}
\hline
\multicolumn{1}{c}{\multirow{2}{*}{Pr.}} &  & \multicolumn{2}{c}{Temperature=0}                     &  & \multicolumn{2}{c}{Temperature=0.5}                   &  & \multicolumn{2}{c}{Temperature=1.0}                   &  & \multicolumn{2}{c}{Temperature=1.5}                   &  & \multicolumn{2}{c}{Temperature=2.0}                   &  & \multicolumn{2}{c}{Avg (Tem.$\leq$1.5)}           \\ \cline{3-4} \cline{6-7} \cline{9-10} \cline{12-13} \cline{15-16} \cline{18-19} 
\multicolumn{1}{c}{}                     &  & \multicolumn{1}{c}{EM-T} & \multicolumn{1}{c}{BLEU-T} &  & \multicolumn{1}{c}{EM-T} & \multicolumn{1}{c}{BLEU-T} &  & \multicolumn{1}{c}{EM-T} & \multicolumn{1}{c}{BLEU-T} &  & \multicolumn{1}{c}{EM-T} & \multicolumn{1}{c}{BLEU-T} &  & \multicolumn{1}{c}{EM-T} & \multicolumn{1}{c}{BLEU-T} &  & \multicolumn{1}{c}{EM-T} & \multicolumn{1}{c}{BLEU-T} \\ \hline
P1                                       &  & 19.22 (0.23)                    & 73.58 (0.22)                      &  & 18.30 (0.54)                    & 72.82 (0.53)                      &  & 16.48 (0.77)                    & 71.15 (0.45)                      &  & 12.27 (1.65)                    & 64.62 (0.57)                      &  & 6.49 (0.75)                     & 28.76 (1.21)                      &  & 16.57                    & 70.54                      \\
P2                                       &  & \textbf{21.48} (0.33)           & \textbf{77.49} (0.27)             &  & 19.76 (1.01)                    & 76.40 (0.95)                      &  & 16.66 (0.77)                    & 74.12 (0.29)                      &  & 11.69 (0.71)                    & 65.48 (0.10)                      &  & 3.59 (0.57)                     & 14.82 (0.24)                      &  & 17.40                    & 73.37                      \\
P3                                       &  & 16.40 (0.23)                    & 75.37 (0.17)                      &  & 15.76 (0.27)                    & 74.66 (0.41)                      &  & 13.02 (1.02)                    & 71.92 (1.33)                      &  & 9.06 (0.09)                     & 63.36 (0.88)                      &  & 3.89 (0.25)                     & 21.50 (0.37)                      &  & 13.56                    & 71.33                      \\
P4                                       &  & 19.22 (0.10)                    & 75.30 (0.16)                      &  & 18.62 (0.59)                    & 74.68 (0.42)                      &  & 16.98 (0.36)                    & 72.66 (0.81)                      &  & 11.83 (0.77)                    & 65.62 (0.22)                      &  & 6.39 (0.49)                     & 25.21 (0.93)                      &  & 16.66                    & 72.06                      \\
P5                                       &  & 21.16 (0.44)                    & 76.66 (0.29)                      &  & 19.93 (0.37)                    & 76.35 (0.43)                      &  & 16.89 (0.85)                    & 74.69 (0.78)                      &  & 10.48 (0.50)                    & 63.96 (1.08)                      &  & 1.78 (0.75)                     & 14.25 (0.29)                      &  & 17.11                    & 72.92                      \\ \hline
Avg                                      &  & 19.50                    & 75.68                      &  & 18.47                    & 74.98                      &  & 16.01                    & 72.91                      &  & 11.06                    & 64.61                      &  & 4.43                     & 20.91                      &  & 16.26                    & 72.05                      \\ \hline
\end{tabular}
}
\end{table*}


\subsubsection{Setup.} Prompts and temperatures are two crucial parameters that can significantly impact the performance of ChatGPT in code refinement tasks. To determine the optimal values for these parameters, we conducted an experiment to evaluate their impact on code refinement. \guo{Note that while temperatures and prompts are parameters utilized by ChatGPT, they are not applicable to run CodeReviewer. CodeReviewer solely relies on the concatenation of old code and code reviews as its input.}

Specifically, temperature is a parameter that controls the level of randomness and creativity in the generated output of ChatGPT. Higher temperature settings tend to produce more diverse and innovative responses, but with a higher risk of generating nonsensical or irrelevant output. \guo{In order to explore the effects of different temperature settings in ChatGPT, which ranges from 0 to 2, we chose five specific temperature values (i.e., 0, 0.5, 1.0, 1.5, and 2.0) due to the high cost of ChatGPT API.}

\guo{
To select the prompts, we followed the established best practices ~\cite{promptingguide,chatgpt-prompts} which suggests that prompts could include four types of elements, i.e., \textit{Instruction}, \textit{Context}, \textit{Input Data} and \textit{Output Indicator}. We have tried prompts with various combinations of these four elements. During our preliminary exploration stage, we experimented with a total of 14 prompts. Due to budget constraints, we selected the 5 best-performing and representative prompts:}

\begin{enumerate}[leftmargin=*]
    \item \textbf{Prompt 1 (P1): the simplest prompt.} We only provided the basic requirement of generating new code based on the old code and review, without additional description.
    \item \textbf{Prompt 2 (P2): P1 + Scenario Description.}  P2 was designed based on Prompt 1 but included a scenario description that asked ChatGPT to act as a developer and modify the code based on review information from a pull request that is from the team leader. 
    \item \textbf{Prompt 3 (P3): P1 + Detailed Requirements.} P3 included detailed requirement information, such as keeping the original content and format of the code as much as possible and not completing any code snippets in the old code or modifying any code not mentioned in the review. 
     \item \textbf{Prompt 4 (P4): P1 + Concise Requirements.} Similar to P3,  P4 also included requirement information that is more concise.
     \item  \textbf{Prompt 5 (P5): P4 + Scenario Description.} P5 was a combination of Prompts 2 and 4, containing both scenario description and requirement information. 
\end{enumerate}

\begin{figure}[!t]
\centering
\includegraphics[width=1\linewidth]{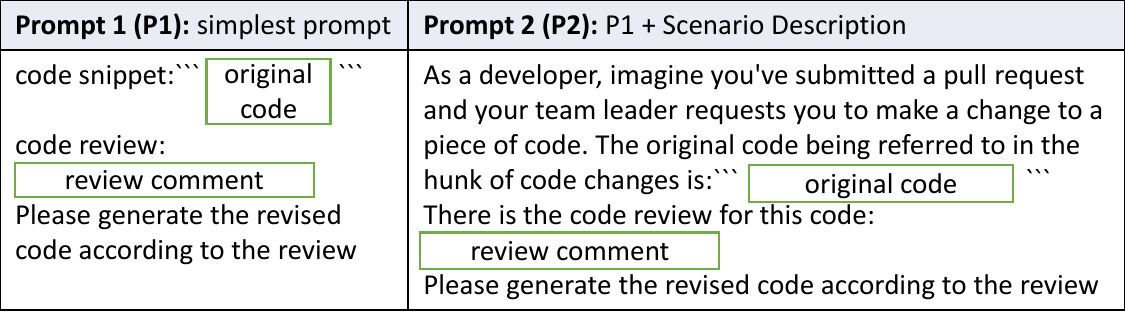}
\caption{\cam{Construction strategies of Prompt 1 and Prompt 2}}
\label{fig:prompts}
\vspace{-3mm}
\end{figure}

\guo{
Specifically, the instruction, context, and output indicator in P1 are all simplest. P2, building upon P1, provides a more detailed context description, while P3, also building upon P1, offers a more detailed output indicator~\cite{chatgptcodereview-settings}. 
\cam{Figure~\ref{fig:prompts} illustrates the construction strategies for Prompt 1 and Prompt 2.} 
The details of the other prompts are available on our website~\cite{ExtraResourceLinkage}. }

To evaluate the effectiveness of ChatGPT under different parameters, we accessed the ChatGPT API and performed code refinement on the CodeReview dataset. Due to the cost of running the ChatGPT API, we randomly selected 500 data entries from the test set of the CodeReview dataset to reduce the number of API calls. To account for the radomness of ChatGPT predictions, we repeated each setting ten times, i.e., making ten ChatGPT API requests on each sample under each setting. We obtained the average of the ten repetitions as the final results.

\subsubsection{Results.}
Table~\ref{tb:settingimpact} displays the results of our evaluation of ChatGPT under different temperature and prompt settings. \guo{Values in parentheses represent standard deviations.} 
Notably, the evaluation results indicate that setting temperature to 0 achieves the best performance for each prompt. As the temperature increases, the performance of ChatGPT decreases significantly. For example, the temperature of 2.0 achieves the worst results.
This phenomenon may be due to the fact that generating new code is a complex and precise task, and high temperature can result in unstable and random results, which are more creative but less reliable. Furthermore, we investigated the results of 500 sampled data with temperature set to 0 with P2, and found that most of the results remain consistent. Specifically, 309 of the data produced the same answers for all 10 runs, while 110 of the data produced only 2 different answers among 10 runs. This finding further underscores the strong stability of using temperature set to 0 for code generation tasks.  Overall, the results suggest that using lower temperature settings tends to produce more stable and better output for code generation tasks. 

\begin{table*}[!t]
\centering
\caption{Impact on trainset and validset.}
\centering
\small
\label{tb:settingimpact_trainset}
\resizebox{.85\textwidth}{!}{
\begin{tabular}{cccccccccccccccc}
\hline
\multicolumn{1}{c}{\multirow{2}{*}{Pr.}} &  & \multicolumn{2}{c}{Temperature=0} &  & \multicolumn{2}{c}{Temperature=0.5} &  & \multicolumn{2}{c}{Temperature=1} &  & \multicolumn{2}{c}{Temperature=1.5} &  & \multicolumn{2}{c}{Temperature=2} \\ \cline{3-4} \cline{6-7} \cline{9-10} \cline{12-13} \cline{15-16} 
\multicolumn{1}{c}{}                     &  & EM-T            & BLEU-T          &  & EM-T             & BLEU-T           &  & EM-T            & BLEU-T          &  & EM-T             & BLEU-T           &  & EM-T           & BLEU-T           \\ \hline
P1                                       &  & 18.1            & 70.77           &  & 18.28            & 70.44            &  & 16.15           & 68.91           &  & 14.08            & 63.21            &  & 2.31           & 6.93             \\
P2                                       &  & 21.55           & 74.21           &  & 20.23            & 73.52            &  & 17.99           & 71.42           &  & 13.45            & 61.94            &  & 1.26           & 3.57             \\
P3                                       &  & 16.21           & 71.2            &  & 16.15            & 71.32            &  & 13.97           & 69.14           &  & 10.4             & 62.87            &  & 1.59           & 4.34             \\
P4                                       &  & 18.28           & 71.45           &  & 17.82            & 71.32            &  & 16.44           & 68.82           &  & 12.36            & 62.48            &  & 1.82           & 5.02             \\
P5                                       &  & 20.11           & 76.17           &  & 19.48            & 75.62            &  & 17.7            & 72.88           &  & 9.94             & 51.69            &  & 0.37           & 2.62             \\ \hline
Avg                                      &  & 18.85           & 72.76           &  & 18.39            & 72.44            &  & 16.45           & 70.23           &  & 12.05            & 60.44            &  & 1.47           & 4.50             \\ \hline
\end{tabular}
}
\end{table*}

Comparing the effects of different prompts under stable temperature settings (0, 0.5, and 1.0), we observed that P2 and P5 achieved significantly better results than others. \guo{Considering the comparative results between P1 and P2, as well as the results between P4 and P5, we can infer that the inclusion of additional scenario descriptions is beneficial in improving the understanding and performance of ChatGPT.}
Furthermore, we noticed that P3 performed worse than P4, despite both prompts containing more requirement information. Sometimes, P3 even performed worse than the simplest prompt, P1. For example, P1 achieved higher EM-trim scores than P3 in all three temperature settings, but P1 was generally worse than P4. This indicates that while providing additional requirement information could be helpful (compared to P1 and P4), too much complex information could harm the performance (P3). It could be because detailed requirement information is more complex to understand by ChatGPT, leading to unstable results.

\cam{To investigate whether the findings of prompts and temperatures also hold across the entire dataset, we conducted an additional experiment. We randomly selected 1,000 data points from the training sets and validation sets of CodeReview dataset, and replicated the experiment. Due to budget constraints, we repeated the experiments for temperatures greater than 1.5 only twice, whereas for other temperature settings, we repeated them 10 times. The results, presented in Table~\ref{tb:settingimpact_trainset}, align closely with the findings in Table 2. Overall, both the EM and BLEU metrics demonstrate comparable performance to that on the test data, further reinforcing the consistent conclusions drawn concerning the influence of temperature and prompt settings as mentioned above.}

\begin{table}[!t]
\vspace{-2mm}
\footnotesize
\caption{Comparisions between Prompt 2 and other prompts.}
\vspace{-2mm}
\label{tb:P2_PVALUE}
\begin{tabular}{ccccc}
\hline
Prompts                        & P1       & P3       & P4       & P5       \\ \hline
EM-T P-value (P2 is superior)   & 4.20E-06 & 7.69E-09 & 2.24E-06 & 0.5320   \\
BLEU-T P-value (P2 is superior) & 9.44E-09 & 2.30E-09 & 1.26E-07 & 0.0039   \\ \hline
\end{tabular}
\vspace{-6mm}
\end{table}

\guo{Table ~\ref{tb:P2_PVALUE} shows the p-value regarding EM-T and BLUE-T between P2 and other prompts with t-test~\cite{ttest_ind}. We can observe that, expect for EM-T P-value (0.5320) between P2 and P5, all p-values are less than 0.005. It implies that P2 significantly outperforms P1, P3, and P4 in terms of both EM-T and BLEU-T scores. As for P5, in terms of EM-T, there is no significant difference between P2 and P5. However, considering the BLEU-T values, P2 is significantly better than P5. Taking into account these factors, we finally selected P2 for conducting the experiments in this paper.}

In the case of unstable temperature settings (1.5 and 2.0), we observed that the overall performance decreased. \guo{Note that, we also tried the fine-grained temperature interval (i.e., 0, 0.1, 0.2, \ldots, 0.9, 1.0) on P2, the results show the similar trend with the larger interval 0.5. The results can be found in the website.}
However, we still noticed that P1 and P4 outperformed other prompts in general. This could be because P1 and P4 are simpler and provide less information, resulting in more stable results under higher temperature settings. In contrast, prompts with more information may make ChatGPT more creative but also more unstable when set with a higher temperature. 


\vskip 1mm
\noindent \fbox{
	\parbox{0.95\linewidth}{\textbf{Answers to RQ1}: The configuration of parameters and temperatures has a significant impact on ChatGPT's performance on code refinement. In most cases, lower temperature settings tend to produce better and more stable results. Prompts involving concise scenario descriptions tend to produce better results.}
}

\subsection{RQ2 Effectiveness of ChatGPT}
Based on the best parameters from RQ1 (i.e., temperature = 0 and prompt 2), we then evaluate ChatGPT on the test dataset of CodeReview (CR) and CodeReview-New (CRN).  Table \ref{tb:quantitative_results} presents the comparative results between ChatGPT and CodeReviewer. The column \#Samples show the number of samples. CodeReview-NewTime (CRNT) and CodeReview-NewLanguage (CRNL) represent the results of two new datasets we constructed (see Table~\ref{tab:codereview_new_sta}), respectively, where CodeReview-NewTime refers to code reviews in the same repositories with code review and CodeReview-NewLanguage refers to code reviews in different repositories with new programming language. \guo{Note that we have also evaluated the performance of ChatGPT on the training and validation datasets of CodeReviewer. The detailed results of these evaluations are available on our website~\cite{ExtraResourceLinkage} due to space limitations. The results demonstrate similar performance to that observed on the test dataset and show the consistent conclusions drawn regarding the impact of temperature and prompt settings in RQ1.} 
We can see that ChatGPT achieves stable results across different datasets. In particular, the evaluation results suggest that ChatGPT performs better on CodeReview-New compared to CodeReview due to the higher quality of reviews in CodeReview-New.

\guo{We further conducted an in-depth analysis to understand the lower performance of CodeReviewer compared to ChatGPT on the new dataset. We identified 2,283 cases from the new dataset where ChatGPT provided a correct response while CodeReviewer did not. We randomly selected 150 of them for the manual analysis. Through our analysis, we identified 4 main root causes:
}
\guo{
\begin{itemize}[leftmargin=*]
    \item \textit{(34) Inaccurate understanding of the review content}. We have observed that some code reviews contain unclear information, such as ambiguous location references, unclear changes, or requiring domain-specific knowledge, which is challenging for the CodeReviewer model to comprehend. 
    \item \textit{(62) Over deletion}. CodeReviewer model exhibits a tendency to inaccurately delete code snippets. Specifically, in 30 cases, the CodeReviewer model erroneously deleted correct code snippets that should have been preserved. Additionally, in 32 cases, the model deleted a significant portion of code snippets that required modifications, resulting in excessive deletions. 
    \item \textit{(10) Extra modification}. In some cases, CodeReviewer model may introduce unnecessary modifications to code snippets that do not require any changes.
    \item \textit{(44) Hard to understand the ground truth provided in the code block.} Our analysis has revealed that, in some cases, reviewers have accurately suggested changes within the code block. However, CodeReviewer fails to recognize that the code within these blocks represents the ground truth, leading to incorrect modifications.
\end{itemize}
}

\guo{
In summary, the main root cause appears to be the different understanding ability of the models. The CodeReviewer model struggles with comprehending some unclear reviews, while ChatGPT demonstrates a stronger ability to capture the underlying semantics accurately. We have included examples that illustrate the root causes and the different performance of the models on our website~\cite{ExtraResourceLinkage}.
}

\begin{table}[!t]
\centering
\caption{Quantitative evaluation results.}
\label{tb:quantitative_results}
\resizebox{.45\textwidth}{!}{
\begin{tabular}{ccccccc}
\hline
                     Dataset                 &       Tool       & \#Samples & EM             & EM-T       & BLEU           & BLEU-T     \\ \hline
\multirow{2}{*}{$CR$}           & CodeReviewer & \multirow{2}{*}{13,104} & \textbf{32.49} & \textbf{32.55} & \textbf{83.39} & \textbf{83.50} \\
                                      & ChatGPT      &   & 16.70          & 19.47          & 68.26          & 75.12          \\ \hline
\multirow{2}{*}{$CRN$}       & CodeReviewer & \multirow{2}{*}{14,568} & 14.84          & 15.50          & 62.25          & 62.88          \\
                                      & ChatGPT      &  & \textbf{19.52} & \textbf{22.78} & \textbf{72.56} & \textbf{76.44} \\ \hline
\multirow{2}{*}{$CRNT$}  & CodeReviewer & \multirow{2}{*}{9,117}  & 15.75          & 16.31          & 62.01          & 62.47          \\
                                      & ChatGPT      &   & \textbf{19.60} & \textbf{22.44} & \textbf{72.90} & \textbf{76.55} \\ \hline
\multirow{2}{*}{$CRNL$} & CodeReviewer & \multirow{2}{*}{5,451}  & 13.21          & 14.05          & 62.67          & 63.61          \\
                                      & ChatGPT      &   & \textbf{19.39} & \textbf{23.40} & \textbf{71.97} & \textbf{76.25} \\ \hline
\end{tabular}
}
\end{table}

Although ChatGPT outperforms CodeReviewer on the new dataset, the results are still not as good as expected, with an EM-trim score of only 22.78. This indicates that ChatGPT still requires significant improvement in code refinement tasks, motivating further exploration of its strengths and weaknesses in RQ3 and RQ4. 

\guo{
Furthermore, our observations indicate that ChatGPT often generates additional text that explains its code refinements. This extra text can offer both advantages and disadvantages. On one hand, it provides explanations that assist users in understanding the code refinements and assessing the reasonableness of the changes made. On the other hand, it may require users to make an additional effort to remove this extra text when submitting the refined code. However, we believe that automatic filtering out such extra text is relatively easier since ChatGPT frequently encloses the code with code blocks, typically denoted by three backticks.
}

\vskip 1mm
\noindent \fbox{
	\parbox{0.95\linewidth}{\textbf{Answers to RQ2}: \guo{Overall, ChatGPT demonstrates better generalization capabilities than CodeReviewer when applied to unseen dataset.} However, its effectiveness is still limited, with EM-trim and BLEU-trim scores of only 22.78 and 76.55, respectively.}
}

\subsection{RQ3 Strengths and Weaknesses of ChatGPT}
\subsubsection{Setup.}
To gain a deeper understanding of the strengths and weaknesses of ChatGPT, we conducted a qualitative analysis on the results of RQ2. Specifically, we randomly selected 400 samples, including 200 samples each from the CodeReview and CodeReview-New datasets, which achieved 90\% confidence level and 5.8\% confidence interval.
Then we manually annotated them along three dimensions: the relevance of the review comment to the code refinement (\textit{Comment Relevance}), the information provided by the review comment (\textit{Comment Information}), and the categories of code changes (\textit{Code Change Category}). Our aim was to identify the strengths and weaknesses of ChatGPT based on these three dimensions.


\begin{figure}[!t]
\centering
\includegraphics[width=0.9\linewidth]{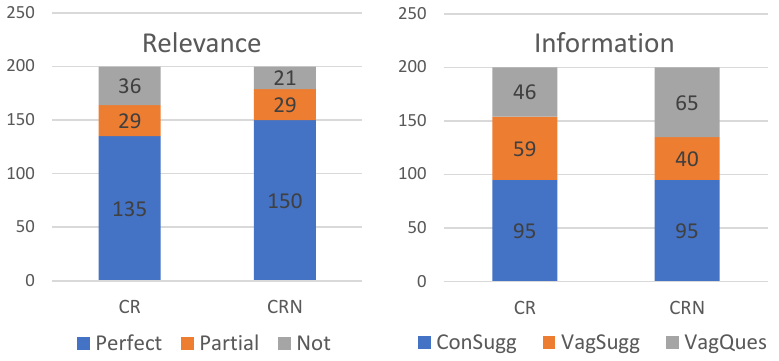}
\caption{Data quality of CodeReview and CodeReview-New.}
\label{fig:data_quality}
\end{figure}

We employed a rigorous annotation process for the manual study of ChatGPT on the selected samples. To facilitate the annotation process, we developed a annotation website that allowed annotators to view the review comment, the original code $C_1$, the ground truth revised code $C_2$, and the original pull request link in a single page. The annotators were able to refer to the code, discussions, and commits in the original pull request if necessary to determine the annotation categories.
Two co-authors independently annotated the samples along the three dimensions. 
When discrepancies occurred between the annotations of the two co-authors, a third author was consulted to resolve the issue through discussion. Conflicts were resolved every 50 samples, and annotation standards were aligned over eight rounds to ensure consistency and accuracy in the annotation process. It took 14 people days to perform the annotation in total. The final Cohen's Kappa coefficient ~\cite{mchugh2012interrater} for Comment Relevance, Comment Information, and Code Change Category was 0.675, 0.696 and 0.888 respectively, suggesting moderate, moderate and strong agreement between the two annotators. 


\textbf{Comment Relevance} measures the degree of relevance between the review comments and the code changes in the test dataset, reflecting the quality of the dataset. 
The relevance of the comments is divided into three levels:

\begin{itemize}[leftmargin=*]
 \item \textbf{Level 1 (Not):} There is no apparent relationship between the code change and the review comment.
    \item \textbf{Level 2 (Partial): } The suggestions in the review comment are partially implemented in the code change, or some refinement in the code change is not present in the suggestions of the comment.
    \item \textbf{Level 3 (Perfect):} The code changes strictly follow the review comment, and there is a clear correspondence between them. In other words, the suggestion of the review comment is fully implemented in the code change, and the code refinement is entirely contained within the review comment.
   
\end{itemize}


\textbf{Comment Information} measures the sufficiency and clarity of the instructions contained in the comment regarding the code change, which reflects the difficulty for the contributor or a model to refine the code.
For example, a comment like ``There are spaces missing'' is more informative than ``This function name does not describe well what it does.''  We followed the definition of comment information from~\cite{ms2022codereviewer}, and divided the comment information into three levels:
\begin{itemize}[leftmargin=*]
    \item \textbf{Level 1 (Vague Question)}: The review comment only gives a general direction for modification (e.g., ``we should maintain the consistency of variable naming'') without clear suggestions for changes.
    \item \textbf{Level 2 (Vague Suggestion)}: The review comment provides specific suggestions for modification (e.g., ``changing it with camel case style''), but does not directly specify the location of the code that should be modified.
\item \textbf{Level 3 (Concrete Suggestion)}: The review comment includes explicit requests for adding or modifying code snippets (e.g., ``changing the variable name 'testfile' to 'testFile''') or explicitly identifies code snippets to be removed.

\end{itemize}



\textbf{Code Change Category} is used to measure the intention of the code changes. We followed the taxonomy in~\cite{tufano2021towards} and defined the categories based on our annotations. There are 4 major categories, including \textit{Documentation Category}, \textit{Feature Category}, \textit{Refactoring Category}, and \textit{Documentation-and-Code Category}.

\begin{itemize}[leftmargin=*]
\item \textbf{Documentation Category} represents code changes that only add, modify, or remove documentation. Modifications according to conventions (Documentation-conventions) may also involve additions, modifications, or deletions, but we separated it for easier analysis of the unique challenges it poses to the model's prediction of revised code.
\item \textbf{Feature Category} represents code changes in terms of functional logic, such as adding, modifying, or removing code.
\item \textbf{Refactoring Category} refers to non-functional code refactoring, including renaming code entities (Refactoring-rename), swapping two code snippets (Refactoring-swap), and updating code based on coding standards (Refactoring-conventions).
\item \textbf{Documentation-and-Code Category} represents code changes that include both documentation and code modifications.
\end{itemize}

\begin{figure}[!t]
\centering
\includegraphics[width=0.8\linewidth]{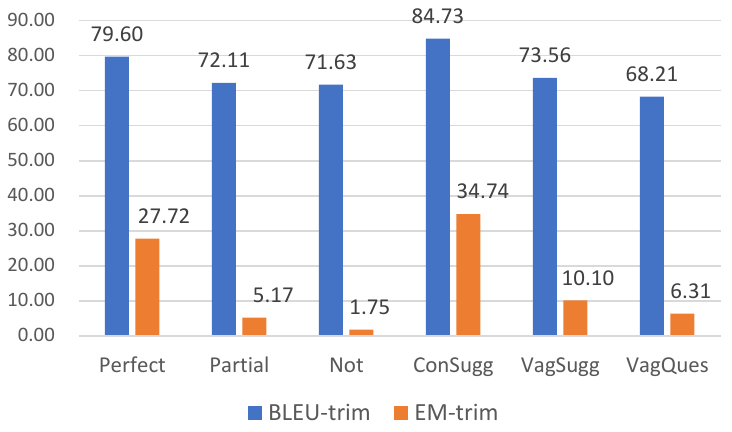}
\caption{Qualitative results of ChatGPT on data with different review information levels.}
\label{fig:relevance_info}
\end{figure}

Figure ~\ref{fig:data_quality} presents the results of the annotation  on the CodeReview dataset and the CodeReview-New dataset, which measures comment relevance and comment information. The results show that, compared to the CodeReview dataset, CodeReview-New dataset, constructed with stricter filtering rules, has more samples with \textit{perfect} relevance levels (150 vs. 135) and fewer samples with \textit{not} relevance levels (21 vs. 36), indicating higher quality. Furthermore, the CodeReview-New dataset has fewer samples with \textit{vague suggestion} level  (40 vs. 59) and more samples with \textit{vague question} level  (65 vs. 46) than the CodeReview dataset. 


Figure \ref{fig:relevance_info} illustrates the results of ChatGPT on different comment relevance and information levels. The figure highlights that ChatGPT performs the best when the comments are classified as \textit{perfect} relevance, outperforming both \textit{partial} and \textit{not} relevance levels. In addition, ChatGPT performs the best on reviews that contain \textit{concrete suggestion} information, while performing similarly for \textit{vague suggestions} and \textit{vague questions}. The results imply that the quality of data significantly impacts ChatGPT's performance, as reviews with low relevance and low information do not provide enough context and information for ChatGPT to make accurate predictions. 


\begin{figure}[!t]
\centering
\includegraphics[width=0.95\linewidth]{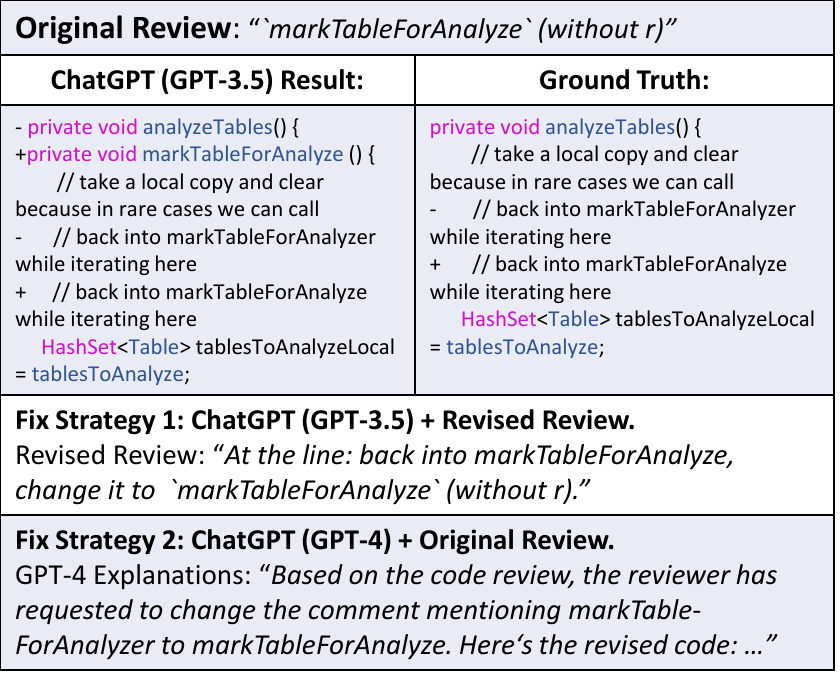}
\caption{An example of unclear location and the mitigation.}
\label{fig:location_example}
\end{figure}

\begin{table*}[!t]
\caption{Results of ChatGPT on different code changes.}
\label{tb:code_change_results}
\centering
\footnotesize
\setlength{\tabcolsep}{1pt}
\resizebox{.85\textwidth}{!}{
\begin{tabular}{cccccccccccc}
\hline
         & Doc-add & Doc-rem & Doc-mod & Doc-con & Feat-add & Feat-rem & Feat-mod & Ref-ren & Ref-swap & Ref-con & Doc\&Code \\ \hline
\#Sample & 14      & 8       & 55      & 13      & 21       & 52       & 153      & 24      & 6        & 34      & 20        \\
EM-T     & 0.00    & 50.00   & 16.36   & 23.08   & 4.76     & 23.08    & 19.61    & 29.17   & 33.33    & 44.12   & 0.00      \\
BLEU-T   & 52.65   & 87.24   & 81.16   & 67.45   & 75.40    & 73.27    & 79.43    & 85.88   & 82.14    & 82.22   & 64.09     \\ \hline
\end{tabular}
}
\end{table*}

Table \ref{tb:code_change_results} summarizes the results across different code change categories. It shows that ChatGPT performs best in the Refactor category with an EM-trim of 37.50\% and a BLEU-trim of 83.58\%, indicating that ChatGPT has a good understanding of how to perform code refactoring. However, the \textit{Documentation-and-Code} category is the weakest performing category, with an EM-trim of 0\% and a BLEU-trim of 64.09\%, which highlights the difficulty in making simultaneous changes to code and documentation while maintaining consistency. When comparing minor categories, ChatGPT is best at handling \textit{remove}-type code changes, followed by \textit{modify} and \textit{add} categories. Additionally, we observed that some of predictions about updates and adds are actually correct, but do not strictly match the ground truth answers, which will be discussed in RQ4.
The results also suggest that ChatGPT is skilled at updating code based on conventions, with EM-trim values of 23.08\% and 44.12\% for Documentation-convention and Refactor-convention samples, respectively, while the average EM-trim for the Documentation and Refactor categories is lower at 17.78\% and 37.50\%, respectively.

\vskip 1mm
\noindent \fbox{
	\parbox{0.95\linewidth}{\textbf{Answers to RQ3}: ChatGPT performs better on high-quality reviews with concrete suggestions, while its performance is worse on reviews with low relevance and low information. Furthermore, ChatGPT demonstrates the highest performance on code refactoring tasks, while its performance is lower on tasks that involve refining documentation and functionalities.}
}

\subsection{RQ4 Root Causes Analysis and Mitigation}
In RQ4, we aim to further understand the root causes of ChatGPT's underperforming cases and identify potential solutions for improvement. 
Specifically, we collect 206 underperforming cases that met two criteria: 1) the reviews have \textit{perfect relevance} and 2) the EM-trim scores calculated based on outputs of ChatGPT were 0. 
\subsubsection{Root Cause Analysis}
Table \ref{tb:causes} presents the results of the root cause analysis, which includes two major categories of root causes: \textit{inaccurate measurement} and \textit{incorrect prediction}. 


\textbf{Inaccurate Measurement Category} refers to false positives where the predicted refinement by ChatGPT is correct based on our manual inspection, but the measurement metrics, such as EM or EM-trim, are low due to the strict matching. Four types of root causes were identified in this category:
\textit{Insignificant Omission (IO)}, where ChatGPT did not return unmodified code segments but correctly returned the modified parts;
\textit{Unexpected Grammar Fix (UGF)}, where ChatGPT fixed grammar errors in the documentation that were not present in the ground truth revised code;
\textit{Code Style Difference (CSD)}, where the predicted code by ChatGPT is semantically identical to the ground truth revised code, with differences only in whitespace, line breaks, and other code style aspects that do not affect code semantics, and the review comment did not explicitly prohibit the change of code style. 
{
\textit{Reasonable Improvement (RI)}, refers to cases where ChatGPT's modifications are highly reasonable and represent an improvement over the original version.
}
 
\begin{figure}[!t]
\centering
\includegraphics[width=0.95\linewidth]{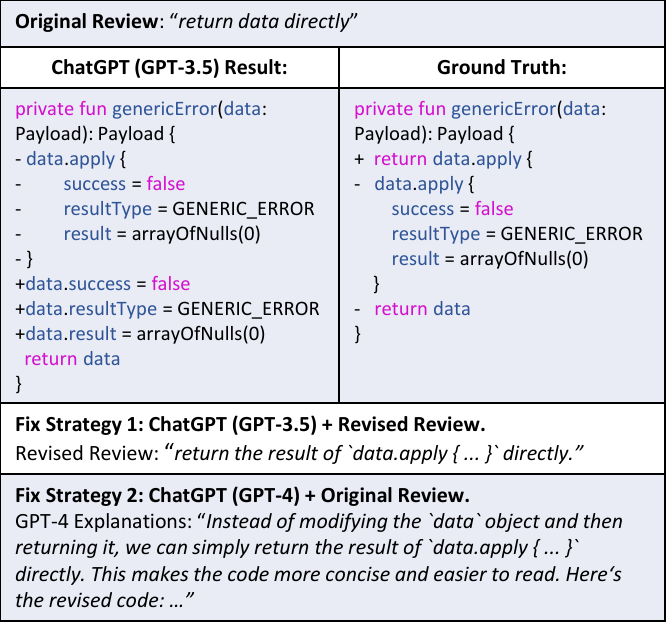}
\caption{\guo{An example of unclear changes and the mitigation.}}
\label{fig:information_example}
\end{figure}

\begin{figure}[!t]
\centering
\includegraphics[width=1\linewidth]{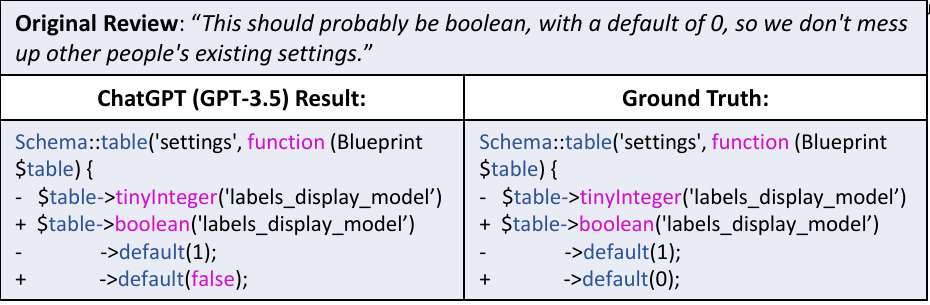}
\caption{\cam{An example of model fallacy.}}
\label{fig:modelfallacy_example}
\end{figure}

\textbf{Incorrect Prediction Category} refers to true positive cases where ChatGPT made incorrect answers compared to the ground truth revised code. We identified three types of root causes in this category. \textit{Need Domain Knowledge (NDK)} refers to cases where the review comment does not provide the necessary repository-related domain knowledge to complete the modification (e.g., ``change this as the style in {\textit{anotherFile}}''). \textit{Unclear Location (UL)} refers to cases where the review comment does not provide a specific location for the code to be modified. For example, in Figure \ref{fig:location_example},  the review does not clearly indicate the location of the changes, and ChatGPT (GPT-3.5) erroneously modifies the function name as well.  Although contributors can see the specific location of the review comment on the GitHub pull request interface, such information is not provided to ChatGPT, following the same settings as CodeReviewer~\cite{ms2022codereviewer}.  \textit{Unclear Changes (UC)} refers to cases where the review comment has a lower information level, causing ChatGPT to be unable to determine the specific modifications needed, resulting in underperformance. 
\guo{For example, in Figure \ref{fig:information_example}, ChatGPT (GPT-3.5) mistakenly assumes that the review suggests returning the result of ``data.apply{...}'' to data itself due to the vague comment.}
\guo{\textit{Model Fallacy (MF)} refers to cases where the review is accurate and clear from the perspective of human, yet ChatGPT fails to handle them correctly. It suggests that the observed issues are more likely to be inherent to the model itself rather than solely stemming from the quality of the review.} 
\cam{As an illustration, in Figure \ref{fig:modelfallacy_example}, ChatGPT (GPT-3.5) mistakenly believes that the review suggests changing default(1) to default(false).}



As presented in Table \ref{tb:causes}, 51 (20.39\%) of the underperforming cases were caused by inaccurate EM measurement. For the remaining 164 (79.61\%) cases where ChatGPT outputs incorrect answers, the majority 107 (51.94\%) cases were caused by the lack of domain knowledge required to complete the modification. Another 44 cases (21.36\%) were due to unclear location information in the review comment, while 13 cases (6.31\%) were caused by unclear instructions provided in the review comments.

\subsubsection{Mitigation Strategies}
We further investigated potential mitigation to improve ChatGPT on the underperforming cases in the \textit{Incorrect Prediction }category as  \textit{Need Domain Knowledge} requires more information. 
\guo{
In general, mitigation can be conducted from two main directions: \textit{ improving the quality of review comments} and \textit{enhancing the models} used for code refinement. Improving the review quality can be achieved through two avenues: \textit{designing best practices for reviewers to provide high-quality reviews} and\textit{ developing tools to assist in refining low-quality reviews} if the reviewers cannot provide high-quality ones. 
In this study, we would like to investigate whether providing more precise reviews and using more advanced models can improve the performance of LLMs on the code refinement task. We leave the study of advanced mitigation strategies (e.g., automatic review refinement) as the future work.}

\begin{table}[!t]
\caption{\guo{Results of root cause analysis.}}
\footnotesize
\label{tb:causes}
\begin{tabular}{cccccccccccc}
\hline
          &  & \multicolumn{4}{c}{Inaccurate Measurement} &  & \multicolumn{4}{c}{Incorrect Prediction} &  \\ \cline{3-6} \cline{8-11}
Type      &  & IO       & UGF       & CSD       & RI      &  & NDK       & UL       & UC      & \guo{MF}      &  \\ \hline
\#Samples &  & 13       & 2         & 19        & 8       &  & 107       & 32       & 11      & \guo{14}      &  \\ \hline
\end{tabular}
\end{table}

For the cases related to \textit{Unclear Location} and \textit{Unclear Changes},
\guo{we identified three strategies for improving the quality of reviews and models: incorporating specific location information in the review (abbreviated as Loc.), providing more explicit review comments (abbreviated as Exp.), and using more advanced GPT-4 model in ChatGPT. }
When utilizing GPT-4, in addition to employing the original review directly (abbreviated as Dir.), we can also add specific location information or provide more explicit review comments if needed.
\guo{We aim to study whether the strategies could mitigate these challenges of ChatGPT.}

Table \ref{tb:mitigation} shows the results with different mitigation strategies. The rows UL and UC refer to the cases under \textit{Unclear Location} and \textit{Unclear Changes}, respectively. The results show that GPT-3.5, combined with the corresponding mitigation techniques, can resolve 24/32 (75\%) of \textit{Unclear Location} cases and 6/11 (54.54\%) of \textit{Unclear Changes} cases. By simply switching to GPT-4 without using mitigation techniques, it can resolve cases very close to those addressed by GPT-3.5 with mitigation techniques. After applying the mitigation techniques, GPT-4 can resolve 31/32 (96.88\%) of Unclear Location and 10/11 (90.91\%) of Unclear Changes cases.
Figure \ref{fig:location_example} and Figure \ref{fig:information_example} show two examples with different mitigations. By revising the original review (i.e., adding location information and making it more explicit), ChatGPT (GPT-3.5) can accurately refine the code. Another method is to use a more advanced LLM, i.e., GPT-4, which is capable of directly producing correct results without the need for review revision. 
In addition, we show part of explanations generated by GPT-4, which are clear and reasonable. Moreover, unlike GPT-3.5, GPT-4 often asks the reviewer for specific modification locations or content when it cannot infer them from the review comment. This is particularly useful when applied in real-world scenarios, as it allows for iteratively helping the reviewer refine their review comment until the model can better understand it, ultimately improving the accuracy of the predicted code changes.

\vskip 1mm
\noindent \fbox{
	\parbox{0.95\linewidth}{\textbf{Answers to RQ4}: The main root causes identified in our analysis were the lack of domain knowledge, unclear location, and unclear changes. Two potential directions for mitigating these issues were identified: improving the large language model, such as using GPT-4 instead of GPT-3.5, and improving the quality of reviews, such as providing more clear information.}
}
\begin{table}[!t] 
\caption{Results of mitigation strategies.}
\label{tb:mitigation}
\resizebox{.45\textwidth}{!}{
    \begin{tabular}{cccccccccccc}
    \hline
    \multirow{2}{*}{Strategy}  &  \multirow{2}{*}{\#Samples} &  & \multicolumn{3}{c}{GPT-3.5} &  & \multicolumn{4}{c}{GPT-4}  &  \\ \cline{4-6} \cline{8-11}
     &  &  & Loc.    & Exp.    & Total   &  & Dir. & Loc. & Exp. & Total &  \\ \hline
    UL       & 32        &  & 24      & -       & 24      &  & 22   & 9   & -    & 31    &  \\
    UC       & 11        &  & -       & 6       & 6       &  & 6    & -    & 4    & 10    &  \\ \hline
    \end{tabular}

}
\end{table}

\section{Implications}

Our study provides implications for both developers seeking to automate code refinement and researchers working in the code review field.

\textbf{Developers:} 
 Our findings show that ChatGPT has the potential to significantly aid developers in code refinement tasks. However, the results also suggest that developers must configure language models like ChatGPT carefully, ensure review quality, and validate output. Our study highlights the impact of temperature and prompt configuration on performance, suggesting that using lower temperatures and concise descriptions with scenario information can lead to better and more stable results. Developers should therefore carefully configure these parameters before using LLMs for code refinement tasks. 
 \guo{Regarding the reviewers who create the code reviews, we have found that clearer reviews significantly aid ChatGPT in understanding modification suggestions. We suggest reviewers to write more specific and detailed review comments. Specifically, the reviewers should be careful in using specific syntax (e.g., code blocks) that may be difficult to be understood by ChatGPT. A safe solution could be that the reviewers can check the clarity of the review content with ChatGPT. For developers who utilize ChatGPT for automated code modification, we recommend conducting a careful manual review of ChatGPT's results. Especially for modifications requiring strong domain knowledge or cases where the review information is ambiguous, it is important to verify whether ChatGPT correctly understands the reviewer's intent and to check for any unnecessary modifications or deletions made by ChatGPT. One possible way is to read the ChatGPT's explanation carefully to check whether the model understands the review well. Furthermore, we recommend that users to choose advanced models if possible, such as GPT-4, which offer enhanced understanding capabilities.}

\textbf{Researchers:} 
Our study demonstrates that ChatGPT achieves promising results but still has room for improvement. Specifically, we identify some root causes of the underperformance of ChatGPT and propose some strategies to mitigate these challenges. These findings provide important guidance for future research in improving the performance of LLMs and enhancing the quality of code reviews. Potential research directions include automatic generation of high-quality reviews, review refinement, and automatic low-quality review detection and filtering. 
Furthermore, our study highlights the limitations of existing metrics such as EM and BLEU, suggesting the need for more accurate and reliable metrics for evaluating the results of language models in code refinement tasks. 
\section{Threats To Validity}
The selected baseline model and benchmark could be a threat to the validity of our results. We addressed this by selecting a state-of-the-art method as reference and creating a new test dataset, $CRN$, with stricter filtering rules.
The randomness of ChatGPT predictions is another potential threat to the validity of our results. To mitigate this, we ran each setting ten times in RQ1, which provided us with more reliable and stable results. In RQ2, we did not run multiple times due to the high cost of accessing ChatGPT API.
The prompts settings we used for ChatGPT could be a threat, as there may be other optimal prompts for code refinement tasks. \guo{Moreover, the different wording of the prompts could also impact the results. We try to address this by following the existing best practices and selecting a range of prompts with varying levels of complexity and specificity, which allowed us to study which types of prompts worked best in different contexts. Another potential threat arises from the comparison between ChatGPT and CodeReviewer, which involve different settings. Specifically, in RQ1, we empirically determined the optimal parameters for temperature and prompts in ChatGPT. We assume that CodeReviewer also achieves its best performance with its hyper-parameter settings.}

The randomness of the selection of samples for the manual annotation process could also be a threat. However, we believe that this would not affect the overall conclusions drawn from our results, especially on the performance of ChatGPT on different categories in RQ3. The subjective nature of human decisions in the manual annotation process is another potential threat to the validity of our results. To address this, we obeyed a rigorous annotation process with two co-authors independently annotating each sample and a third author resolving any inconsistencies or conflicts through discussion. Moreover, the final Cohen's Kappa coefficient indicates relatively high agreement between the two annotators.
\section{RELATED WORK}
\noindent \textbf{Pre-trained Models for SE:}
Large-scale pre-trained models has revolutionized the field of natural language processing ~\cite{radford2018improving,chen2021evaluating}, and its application in the software engineering domain has shown promising results ~\cite{jiang2021cure,kim2021code,aye2021learning}. 
Currently, pre-trained model architectures are mainly divided into encoder-only, decoder-only, and encoder-decoder models ~\cite{jiang2021treebert,guo2022unixcoder}.

Encoder-only models pre-train a bidirectional Transformer, which can access token information before and after the current token when training~\cite{feng2020codebert, guo2020graphcodebert, liu2023contrabert}. 
Decoder-only models allow the model to access only the tokens preceding the current token during the training process ~\cite{svyatkovskiy2020intellicode, lu2021codexglue}. 
GPT-3~\cite{gpt3} also employs decoder-only architectures and has a significantly larger parameter size (175 billion, 10x more than any previous LLMs). 
Additionally, GPT-3.5-Turbo ~\cite{gpt35turbo}, the default model of ChatGPT, adopt Reinforced Learning with Human Feedback (RLHF) to enhance GPT3's ability to understand instructions and generate content aligned with human expectations.
CodeT5 ~\cite{codet5} is a typical pretraining model for code utilizing an encoder-decoder architecture. It adopts the T5 ~\cite{t5} model and considers crucial token type information from identifiers during pretraining. CommitBART ~\cite{liu2023commitbart} also employs an encoder-decoder architecture and is specially trained for commit representation. There are also some works focusing on exploring the learned program semantics for these pre-trained models in SE~\cite{ma2023scope, ma2022self} and analyzing the robustness~\cite{liu2023contrabert} and security~\cite{li2023multi} of these models.

\noindent \textbf{Automating Code Review Activities:}
Studies have presented evidence that developer spend a considerable amount of time on code review activities ~\cite{paixao2020behind,bosu2015chatgpt-promptsracteristics}, both writing review comments for other's code and performing code changes according to other's comments \cite{mastropaolo2021studying,bosu2013impact}. 
Consequently, numerous studies ~\cite{thongtanunam2015should,zanjani2015automatically,hellendoorn2021towards} have been carried out on automating the code review (ACR) activities, emphasizing their significance and potential impact ~\cite{watson2022systematic}.

According to the stages of code review, prior studies on ACR can be categorized into three tasks ~\cite{zhou2023generation,tufano2019learning,tufano2021towards}: (1) \textit{Code Change Recommendation~\cite{ms2022codereviewer}}: Before the contributor submits the original code for review, the ACR model provides potential code changes that the reviewer might suggest. (2) \textit{Review Comment Generation~\cite{tufano2022using}}: After the contributor submits the code for review, the model provides possible review comments for the reviewer, serving as a draft for review comments. (3) \textit{Code Refinement ~\cite{ms2022codereviewer,tufano2022using}}: After the reviewer provides review comments, the model suggests code changes for the contributor by considering both the review comments and submitted code.
In this paper, we focus on the last task, \textit{Code Refinement}, as it is the final and most crucial step in code review activities. 

Tufano et al.~\cite{tufano2019learning} introduced a Recurrent Neural Network (RNN) based Neural Machine Translation (NMT) model for the code refinement task. 
CodeReviewer~\cite{ms2022codereviewer} utilized the CodeT5 model and designed four pre-training tasks related to code review.
Recently, Zhou et al.~\cite{zhou2023generation} compared existing ACR techniques, including Trans-Review~\cite{tufano2021towards}, AutoTransform~\cite{thongtanunam2022autotransform}, and T5-Review~\cite{tufano2022using}.
They discovered that CodeT5 outperformed existing ACR techniques in both code change recommendation and code refinement tasks.

Although, they evaluated large language models for code, such as CodeT5 and CodeBERT, ChatGPT is significantly different from these LLMs with RLHF and emergent abilities due to a much larger number of parameters ~\cite{ChatGPTblog}, thus need further evaluation.
Despite that ChatGPT have been evaluated on numerous NLP tasks~\cite{hu2023empirical}
and several software engineering tasks ~\cite{sobania2023analysis, white2023chatgpt}, 
this paper presents the first comprehensive empirical study exploring ChatGPT's capabilities in the code refinement task, to the best of our knowledge.

\section{CONCLUSION}
In this paper, we conduct an empirical study to investigate the potential of ChatGPT in automating code review tasks, with a focus on code refinement based on code reviews. We assess the impact of various ChatGPT configurations and examine its effectiveness on both standard code review benchmarks and a new dataset collected by us. Our findings highlight the promising potential of ChatGPT for code refinement, unveil the root causes of its underperformance, and suggest potential strategies to overcome these challenges.

\section{Acknowledgment}
This work was partially supported by the National Key R\&D Project (2021YFF1201102), the National Key R\&D Program of China (2021ZD 0112903), the National Natural Science Foundation of China (Grant No. 61872262), the National Research Foundation, Singapore, and the Cyber Security Agency under its National Cybersecurity R\&D Programme (NCRP25-P04-TAICeN). Any opinions, findings and conclusions or recommendations expressed in this material are those of the author(s) and do not reflect the views of National Research Foundation, Singapore and Cyber Security Agency of Singapore.
\newpage

\bibliographystyle{ACM-Reference-Format}
\bibliography{src/reference}

\end{document}